\documentclass[10pt]{article}
\usepackage{graphicx}
\usepackage{amsmath}
\usepackage{amssymb}
\usepackage{caption2}
\begin{document}
\begin{center}
{\large {\bf \sc{Decipher the width of the $X(3872)$  via the QCD sum rules  }}} \\[2mm]
Zhi-Gang  Wang \footnote{E-mail: zgwang@aliyun.com.  }     \\
 Department of Physics, North China Electric Power University, Baoding 071003, P. R. China
\end{center}

\begin{abstract}
In this work, we take the $X(3872)$ as the hidden-charm tetraquark state with both isospin $I=0$ and $1$ components, then investigate the strong decays $X(3872)\to J/\psi \pi^+\pi^-$, $J/\psi\omega$,  $\chi_{c1}\pi^0$, $D^{*0}\bar{D}^0$ and $D^{0}\bar{D}^0\pi^0$ with the QCD sum rules. We take account of all the Feynman diagrams, and acquire four QCD sum rules based on rigorous quark-hadron duality. We obtain the total decay width about $1\,\rm{MeV}$, which is in excellent agreement with the experiment data $ \Gamma_{X}=1.19\pm 0.21\,\rm{MeV}$  from the Particle Data Group, it is  the first time  to reproduce the tiny  width of the $X(3872)$ via the QCD sum rules, which supports assigning the $X(3872)$ as the hidden-charm tetraquark state with the $J^{PC}=1^{++}$.
 \end{abstract}



\section{Introduction}

In 2003, the Belle collaboration observed  a narrow charmonium-like state, the $X(3872)$, in the $\pi^+\pi^- J/\psi$ mass spectrum in the exclusive
 process $B^\pm \to K^\pm \pi^+\pi^- J/\psi$, which has a mass  of $3872.0\pm 0.6\pm 0.5\,\rm{MeV}$ and a width  less than $2.3\,\rm{MeV}$, the $X(3872)$ lies very near the $D^*\bar{D}$ threshold \cite{Belle-X3872-2003}, which stimulated the interpretation in terms of the $D^*\bar{D}$ molecular state \cite{X3872-mole-Tornqvist,X3872-mole-Swanson,X3872-mole-Fleming,X3872-mole-Liu,X3872-mole-Oset,X3872-mole-Guo,
 X3872-mole-WZG-HT,X3872-mole-WZG-IJMPA,X3872-mole-Mutuk}.  At the same time, other interpretations were suggested, such as the tetraquark state \cite{X3872-Tetra-Maiani,X3872-Tetra-Ebert,X3872-Tetra-Narison-mass,X3872-Tetra-Nielsen-decay,
X3872-Tetra-WZG-HT,WZG-PRD-cc-spectrum}, hybrid state \cite{X3872-hybrid-Li}, charmonium-molecule mixing state \cite{X3872-Charmon-mole-Kalashnikova,X3872-Charmon-mole-Chao,X3872-Charmon-mole-Nielsen,X3872-Charmon-mole-Kang},
 charmonium state \cite{X3872-Charmon-Godfrey}, etc. In fact, the observations of the $X(3872)$ in the $pp$ and $p\bar{p}$ collisions by the CDF, ATLAS, LHCb and CMS collaborations disfavor the pure molecule assignment \cite{X3872-collision-CDF,X3872-collision-ATLAS,X3872-collision-LHCb,X3872-collision-CMS,
 X3872-collision-Grinstein}.

In 2015, the LHCb collaboration studied the angular correlations in the $B^+\to X(3872)K^+$ decays with the subprocess  $X(3872)\to \rho^0 J/\psi \to \pi^+\pi^-\mu^+\mu^-$ to measure orbital angular momentum contributions and to determine the $J^{PC}$  of the $X(3872)$ to be $1^{++}$ \cite{X3872-JPC}. The $X(3872)$ state is probably the best known (and most enigmatic) representative  of the $X$, $Y$ and $Z$ states. One important discriminant between different models is the width of the $X(3872)$. In 2020, the  LHCb collaboration updated the mass and width of the $X(3872)$, and obtained the Breit-Wigner width   $\Gamma=0.96^{+0.19}_{-0.18}\pm 0.21 \,\rm{MeV}$ \cite{X3872-width-JHEP}, or
$1.39 \pm 0.24 \pm 0.10\,\rm{MeV}$ \cite{X3872-width-PRD}, which indicates non-zero width  of the $X(3872)$ and leads to the average width $ \Gamma_{X}=1.19\pm 0.21\,\rm{MeV}$ listed in  "The Review of Particle Physics" \cite{PDG}. As known, we cannot assign a hadron with the mass alone, and we should study the decays to obtain more robust interpretation. Up to today, only the decays $X(3872)\to J/\psi \pi^+\pi^-$, $J/\psi\omega$, $J/\psi\gamma$, $\psi^\prime\gamma$, $\chi_{c1}\pi^0$, $D^{*0}\bar{D}^0$ and $D^{0}\bar{D}^0\pi^0$ are established \cite{PDG}.

In the present work, we will focus on the scenario of tetraquark states.
In Ref.\cite{WZG-PRD-cc-spectrum},  we take the pseudoscalar, scalar, axialvector, vector, tensor (anti)diquarks as the basic constituents, and construct the scalar, axialvector and tensor tetraquark currents to study the mass spectrum of the ground state hidden-charm tetraquark states with the QCD sum rules in a comprehensive way, and observe that the $X(3872)$ can be assigned to be the hidden-charm tetraquark state with the quantum numbers $J^{PC}=1^{++}$. According to the recent combined data analysis, the decays $X(3872)\to J/\psi \rho\to J/\psi \pi^+\pi^-$ and $X(3872)\to J/\psi \omega\to J/\psi \pi^+\pi^-\pi^0$ have almost the same branching fractions \cite{X3872-YuanCZ},  the isospin breaking effects in the decays are large enough and beyond the naive $\rho-\omega$ mixing. In this work, we introduce the isospin breaking effects explicitly and study the decays $X(3872)\to J/\psi \pi^+\pi^-$, $J/\psi\omega$,  $\chi_{c1}\pi^0$, $D^{*0}\bar{D}^0$ and $D^{0}\bar{D}^0\pi^0$ with the QCD sum rules based on rigorous quark-hadron  duality, and try to decipher the width of the $X(3872)$.

The article is arranged as follows:  we obtain  the QCD sum rules for the  hadronic coupling constants in section 2; in section 3, we present numerical results and discussions; section 4 is reserved for our conclusion.

\section{QCD sum rules for  the  hadronic coupling constants}
We write down  the three-point correlation functions  $\Pi_{\mu\nu\alpha}^{1/2}(p,q)$  and $\Pi_{\mu\alpha}^{1/2}(p,q)$  in the  QCD sum rules,
\begin{eqnarray}
\Pi^1_{\mu\nu\alpha}(p,q)&=&i^2\int d^4xd^4y \, e^{ip\cdot x}e^{iq\cdot y}\, \langle 0|T\left\{J_{\mu}^{J/\psi}(x)J_\nu^{\rho}(y)J^{X}_{\alpha}{}^\dagger(0)\right\}|0\rangle\, , \nonumber\\
\Pi^2_{\mu\nu\alpha}(p,q)&=&i^2\int d^4xd^4y \, e^{ip\cdot x}e^{iq\cdot y}\, \langle 0|T\left\{J_{\mu}^{J/\psi}(x)J_\nu^{\omega}(y)J^{X}_{\alpha}{}^\dagger(0)\right\}|0\rangle\, , \nonumber\\
\Pi^1_{\mu\alpha}(p,q)&=&i^2\int d^4xd^4y \, e^{ip\cdot x}e^{iq\cdot y}\, \langle 0|T\left\{J_{\mu}^{\chi}(x)J_5^{\pi}(y)J^{X}_{\alpha}{}^\dagger(0)\right\}|0\rangle\, , \nonumber\\
\Pi^2_{\mu\alpha}(p,q)&=&i^2\int d^4xd^4y \, e^{ip\cdot x}e^{iq\cdot y}\, \langle 0|T\left\{J_{\mu}^{D^*}(x)J_5^{D}(y)J^{X}_{\alpha}{}^\dagger(0)\right\}|0\rangle\, ,
\end{eqnarray}
where the currents

\begin{eqnarray}
J_{\mu}^{J/\psi}(x)&=&\bar{c}(x)\gamma_{\mu} c(x) \, ,\nonumber \\
J_{\mu}^{\chi}(x)&=&\bar{c}(x)\gamma_{\mu}\gamma_5 c(x) \, ,\nonumber \\
J_{\mu}^{D^{*}}(x)&=&\bar{u}(x)\gamma_{\mu} c(x) \, ,
\end{eqnarray}
\begin{eqnarray}
J_{\nu}^{\rho}(y)&=&\frac{1}{\sqrt{2}}\left[\bar{u}(y)\gamma_{\nu} u(y)-\bar{d}(y)\gamma_{\nu} d(y)\right] \, ,\nonumber \\
J_{\nu}^{\omega}(y)&=&\frac{1}{\sqrt{2}}\left[\bar{u}(y)\gamma_{\nu} u(y)+\bar{d}(y)\gamma_{\nu} d(y)\right] \, ,\nonumber \\
J_5^{\pi}(y)&=&\frac{1}{\sqrt{2}}\left[\bar{u}(y)i\gamma_{5} u(y)-\bar{d}(y)i\gamma_{5} d(y)\right] \, ,\nonumber \\
J_5^{D}(y)&=&\bar{c}(y)i\gamma_{5} u(y) \, ,
\end{eqnarray}

\begin{eqnarray}\label{current-JX}
J^X_\alpha(0)&=&\cos\theta J^{u\bar{u}}_{\alpha}(0)+\sin\theta J^{d\bar{d}}_{\alpha}(0)\, ,
\end{eqnarray}

\begin{eqnarray}
J^{u\bar{u}}_{\alpha}(0)&=&\frac{\varepsilon^{ijk}\varepsilon^{imn}}{\sqrt{2}}\Big[u^{T}_j(0)C\gamma_5c_k(0) \bar{u}_m(0)\gamma_\alpha C \bar{c}^{T}_n(0)-u^{T}_j(0)C\gamma_\alpha c_k(0)\bar{u}_m(0)\gamma_5C \bar{c}^{T}_n(0) \Big] \, , \nonumber\\
J^{d\bar{d}}_{\mu}(0)&=&\frac{\varepsilon^{ijk}\varepsilon^{imn}}{\sqrt{2}}\Big[d^{T}_j(0)C\gamma_5c_k(0) \bar{d}_m(0)\gamma_\alpha C \bar{c}^{T}_n(0)-d^{T}_j(0)C\gamma_\alpha c_k(0)\bar{d}_m(0)\gamma_5C \bar{c}^{T}_n(0) \Big] \, ,
\end{eqnarray}
interpolate the mesons $J/\psi$,  $\chi_{c1}$, $D^*$, $\rho$, $\omega$, $\pi$, $D$ and $X(3872)$, respectively.
As the decays $X(3872)\to J/\psi \rho\to J/\psi \pi^+\pi^-$ and $X(3872)\to J/\psi \omega\to J/\psi \pi^+\pi^-\pi^0$ have almost the same branching fractions \cite{X3872-YuanCZ}, which is beyond the naive expectation of the $\rho-\omega$ mixing, we have to introduce mixing effects in the $X(3872)$ and abandon the obsession that the $X(3872)$ has definite isospin $I=0$. We determine the mixing angle $\theta$  by the experimental data via trial and error.

At the hadron side, we insert  a complete set of intermediate hadronic states with the same quantum numbers  as the currents into the three-point correlation functions, and  isolate the ground state contributions explicitly,
\begin{eqnarray}\label{Hadron-CT-1}
\Pi^1_{\mu\nu\alpha}(p,q)&=& \frac{\lambda_X\, f_{J/\psi} m_{J/\psi}\,f_{\rho}m_{\rho}\,G^\prime_{XJ/\psi\rho} }{(m_{X}^2-p^{\prime2})(m_{J/\psi}^2-p^2)(m_{\rho}^2-q^2)}i\varepsilon_{\mu\nu\alpha\sigma}q^\sigma p\cdot q  + \cdots\, , \nonumber\\
&=&\Pi_{\rho}(p^{\prime2},p^2,q^2)\,i\varepsilon_{\mu\nu\alpha\sigma}q^\sigma p\cdot q  + \cdots\, ,
\end{eqnarray}
\begin{eqnarray}\label{Hadron-CT-2}
\Pi^2_{\mu\nu\alpha}(p,q)&=& \frac{\lambda_X\, f_{J/\psi} m_{J/\psi}\,f_{\omega}m_{\omega}\,G^\prime_{XJ/\psi\omega} }{(m_{X}^2-p^{\prime2})(m_{J/\psi}^2-p^2)(m_{\omega}^2-q^2)}i\varepsilon_{\mu\nu\alpha\sigma}q^\sigma p\cdot q  + \cdots\, , \nonumber\\
&=&\Pi_{\omega}(p^{\prime2},p^2,q^2)\,i\varepsilon_{\mu\nu\alpha\sigma}q^\sigma p\cdot q  + \cdots\, ,
\end{eqnarray}

\begin{eqnarray}\label{Hadron-CT-3}
\Pi^1_{\mu\alpha}(p,q)&=& \frac{\lambda_X\, f_{\chi_{c1}} m_{\chi_{c1}}\mu_{\pi}\, G^\prime_{X\chi\pi} }{(m_{X}^2-p^{\prime2})(m_{\chi_{c1}}^2-p^2)(m_{\pi}^2-q^2)}\varepsilon_{\alpha\mu\nu\sigma}p^\nu q^\sigma    + \cdots\, , \nonumber\\
&=&\Pi_{\pi}(p^{\prime2},p^2,q^2)\,\varepsilon_{\alpha\mu\nu\sigma}p^\nu q^\sigma   + \cdots\, ,
\end{eqnarray}
\begin{eqnarray}\label{Hadron-CT-4}
\Pi^2_{\mu\alpha}(p,q)&=& \frac{\lambda_X\, f_{D^*} m_{D^*}\mu_{D}\, G^\prime_{XD^*D} }{(m_{X}^2-p^{\prime2})(m_{D^*}^2-p^2)(m_{D}^2-q^2)}ip\cdot q g_{\mu\alpha}    + \cdots\, ,\nonumber\\
&=&\Pi_{D}(p^{\prime2},p^2,q^2)\,ip\cdot q g_{\mu\alpha}   + \cdots\, ,
\end{eqnarray}
where $p^\prime=p+q$,  $\mu_\pi=\frac{f_\pi m_\pi^2}{m_u+m_d}$, $\mu_D=\frac{f_D m_D^2}{m_c}$,  the hadronic coupling constants $G^\prime_{XJ/\psi\rho}$, $G^\prime_{XJ/\psi\omega}$, $G^\prime_{X\chi\pi}$
and $G^\prime_{XD^*D}$ are defined by
\begin{eqnarray}
\langle J/\psi(p)\rho(q)|X(p^\prime)\rangle&=&-\varepsilon^{\sigma\alpha\mu\nu}p^\prime_\sigma\, \zeta_\alpha\,
\xi^*_\mu\,\varsigma^*_\nu\,p\cdot q\, G^\prime_{XJ/\psi\rho} \, , \nonumber\\
\langle J/\psi(p)\omega(q)|X(p^\prime)\rangle&=&-\varepsilon^{\sigma\alpha\mu\nu}p^\prime_\sigma\, \zeta_\alpha\,  \xi^*_\mu\,\varsigma^*_\nu\,p\cdot q\, G^\prime_{XJ/\psi\omega}  \, , \nonumber\\
\langle \chi_{c1}(p)\pi(q)|X(p^\prime)\rangle&=&i\varepsilon^{\sigma\alpha\mu\nu}p^\prime_\sigma\, \zeta_\alpha\,  p_\mu\,\xi^*_\nu\, G^\prime_{X\chi\pi}  \, , \nonumber\\
\langle D^*(p)D(q)|X(p^\prime)\rangle&=&- \zeta \cdot \xi^* \, p \cdot q\, G^\prime_{XD^*D}  \, ,
\end{eqnarray}
 the  decay constants $\lambda_X$, $f_{J/\psi}$, $f_{\chi_{c1}}$, $f_{\rho}$, $f_{\omega}$, $f_{D^*}$, $f_D$ and $f_{\pi}$, are defined by,
\begin{eqnarray}
\langle0|J_{\mu}^{X}(0)|X(p^\prime)\rangle&=&\lambda_{X} \zeta_{\mu} \,\, ,
\end{eqnarray}
\begin{eqnarray}
\langle0|J_{\mu}^{J/\psi}(0)|J/\psi(p)\rangle&=&f_{J/\psi} m_{J/\psi} \xi_\mu \,\, , \nonumber \\
\langle0|J_{\mu}^{\chi}(0)|\chi_{c1}(p)\rangle&=&f_{\chi_{c1}} m_{\chi_{c1}} \xi_\mu \,\, , \nonumber \\
\langle0|J_{\mu}^{D^*}(0)|D^*(p)\rangle&=&f_{D^*} m_{D^*} \xi_\mu \,\, ,
\end{eqnarray}
\begin{eqnarray}
\langle0|J_{\mu}^{\rho}(0)|\rho(q)\rangle&=&f_{\rho} m_{\rho} \varsigma_\mu \,\, , \nonumber \\
\langle0|J_{\mu}^{\omega}(0)|\omega(q)\rangle&=&f_{\omega} m_{\omega} \varsigma_\mu \,\, ,
\end{eqnarray}
\begin{eqnarray}
\langle0|J_5^{D}(0)|D(q)\rangle&=&\frac{f_{D} m_{D}^2}{m_c}  \,\, ,\nonumber \\
\langle0|J_5^{\pi}(0)|\pi(q)\rangle&=&\frac{f_{\pi} m_{\pi}^2}{m_u+m_d}  \,\, ,
\end{eqnarray}
 the $\zeta_{\mu}$, $\xi_\mu$ and $\varsigma_\mu$  are polarization vectors of the axialvector or vector mesons.

We choose the components $\Pi_{\rho}(p^{\prime2},p^2,q^2)$, $\Pi_{\omega}(p^{\prime2},p^2,q^2)$, $\Pi_{\pi}(p^{\prime2},p^2,q^2)$ and $\Pi_{D}(p^{\prime2},p^2,q^2)$ to study the hadronic coupling constants $G^\prime_{XJ/\psi\rho}$, $G^\prime_{XJ/\psi\omega}$, $G^\prime_{X\chi\pi}$ and $G^\prime_{XD^*D}$, respectively. In Ref.\cite{X3872-Tetra-Nielsen-decay}, the tensor structures $p_\mu \varepsilon_{\nu\alpha\sigma\tau}p^\sigma q^\tau$ and $\varepsilon_{\mu\nu\alpha\sigma}q^\sigma$, which differ from the  structure $\varepsilon_{\mu\nu\alpha\sigma}q^\sigma p\cdot q $ in Eqs.\eqref{Hadron-CT-1}-\eqref{Hadron-CT-2} greatly,  are  chosen to study the hadronic coupling constants $G_{XJ/\psi \rho}$ and $G_{XJ/\psi \omega}$.
The  $G_{XJ/\psi \rho}$ and $G_{XJ/\psi \omega}$ defined in Ref.\cite{X3872-Tetra-Nielsen-decay} (also Ref.\cite{Nielsen-review}) and the  $G^\prime_{XJ/\psi\rho}$ and $G^\prime_{XJ/\psi\omega}$ defined in this work have the relations,
  $G_{XJ/\psi \rho}=p\cdot q  G^\prime_{XJ/\psi\rho}$ and $G_{XJ/\psi \omega}=p \cdot q G^\prime_{XJ/\psi\omega}$, although the two definitions are both reasonable, they lead to quite different QCD sum rules. It is not odd that different predictions may be obtained.
Then, we acquire  the hadronic  spectral densities $\rho_H(s^\prime,s,u)$ through triple  dispersion relation,
\begin{eqnarray}
\Pi_{H}(p^{\prime2},p^2,q^2)&=&\int_{\Delta_s^{\prime2}}^\infty ds^{\prime} \int_{\Delta_s^2}^\infty ds \int_{\Delta_u^2}^\infty du \frac{\rho_{H}(s^\prime,s,u)}{(s^\prime-p^{\prime2})(s-p^2)(u-q^2)}\, ,
\end{eqnarray}
where the $\Delta_{s}^{\prime2}$, $\Delta_{s}^{2}$ and
$\Delta_{u}^{2}$ are thresholds, we add the subscript $H$ to stand for  the components $\Pi_{\rho}(p^{\prime2},p^2,q^2)$, $\Pi_{\omega}(p^{\prime2},p^2,q^2)$, $\Pi_{\pi}(p^{\prime2},p^2,q^2)$ and $\Pi_{D}(p^{\prime2},p^2,q^2)$ at the hadron side.

As the operator product expansion is concerned, we calculate the vacuum condensates up to dimension 5, and obtain the QCD spectral densities through double dispersion relation,
\begin{eqnarray}
\Pi_{QCD}(p^{\prime2},p^2,q^2)&=& \int_{\Delta_s^2}^\infty ds \int_{\Delta_u^2}^\infty du \frac{\rho_{QCD}(p^{\prime2},s,u)}{(s-p^2)(u-q^2)}\, ,
\end{eqnarray}
as
\begin{eqnarray}
{\rm lim}_{\epsilon \to 0}\frac{{\rm Im}\,\Pi_{QCD}(s^\prime+i\epsilon,p^2,q^2)}{\pi}&=&0\, .
\end{eqnarray}
In calculations, we neglect the gluon condensates due to their tiny contributions \cite{WZG-ZJX-Zc-Decay,WZG-Y4660-Decay}.
We accomplish   the integral over $ds^\prime$ firstly at the hadron side, then match the hadron side with the QCD side  bellow the continuum thresholds $s_0$ and $u_0$ to obtain rigorous quark-hadron  duality  \cite{WZG-ZJX-Zc-Decay,WZG-Y4660-Decay},
 \begin{eqnarray}\label{duality}
  \int_{\Delta_s^2}^{s_0}ds \int_{\Delta_u^2}^{u_0}du  \left[ \int_{\Delta_{s}^{\prime2}}^{\infty}ds^\prime  \frac{\rho_H(s^\prime,s,u)}{(s^\prime-p^{\prime2})(s-p^2)(u-q^2)} \right] &=&\int_{\Delta_s^2}^{s_{0}}ds \int_{\Delta_u^2}^{u_0}du  \frac{\rho_{QCD}(s,u)}{(s-p^2)(u-q^2)}
  \, .
\end{eqnarray}
In the following, we write down the hadron representation explicitly,
\begin{eqnarray}\label{duality-explict}
\Pi_{\rho}(p^{\prime2},p^2,q^2)&=& \frac{\lambda_X\, f_{J/\psi} m_{J/\psi}\,f_{\rho}m_{\rho}\,G^\prime_{XJ/\psi\rho} }{(m_{X}^2-p^{\prime2})(m_{J/\psi}^2-p^2)(m_{\rho}^2-q^2)} +\frac{C_{\rho}^\prime}{(m_{J/\psi}^2-p^2)(m_{\rho}^2-q^2)}\, , \nonumber\\
\Pi_{\omega}(p^{\prime2},p^2,q^2)&=& \frac{\lambda_X\, f_{J/\psi} m_{J/\psi}\,f_{\omega}m_{\omega}\,G^\prime_{XJ/\psi\omega} }{(m_{X}^2-p^{\prime2})(m_{J/\psi}^2-p^2)(m_{\omega}^2-q^2)} +\frac{C_{\omega}^\prime}{(m_{J/\psi}^2-p^2)(m_{\omega}^2-q^2)}\, , \nonumber\\
\Pi_{\pi}(p^{\prime2},p^2,q^2)&=& \frac{\lambda_X\, f_{\chi_{c1}} m_{\chi_{c1}}\,\mu_{\pi}\,G^\prime_{X\chi\pi} }{(m_{X}^2-p^{\prime2})(m_{\chi_{c1}}^2-p^2)(m_{\pi}^2-q^2)} +\frac{C_{\pi}^\prime}{(m_{\chi_{c1}}^2-p^2)(m_{\pi}^2-q^2)}\, ,  \nonumber\\
\Pi_{D}(p^{\prime2},p^2,q^2)&=& \frac{\lambda_X\, f_{D^*} m_{D^*}\,\mu_{D}\,G^\prime_{XD^*D} }{(m_{X}^2-p^{\prime2})(m_{D^*}^2-p^2)(m_{D}^2-q^2)} +\frac{C_{D}^\prime}{(m_{D^*}^2-p^2)(m_{D}^2-q^2)}\, ,
\end{eqnarray}
where we introduce the parameters $C^\prime_{\rho/\omega/\pi/D}$ to stand for all the contributions concerning  the higher resonances  in the $s^\prime$ channel,
\begin{eqnarray}\label{duality-C}
C^\prime_{\rho}&=&\int_{s^\prime_0}^{\infty}ds^\prime\frac{\tilde{\rho}_{\rho}(s^\prime,m_{J/\psi}^2,m_{\rho}^2)}{
s^\prime-p^{\prime2}}\, , \nonumber\\
C^\prime_{\omega}&=&\int_{s^\prime_0}^{\infty}ds^\prime\frac{\tilde{\rho}_{\omega}(s^\prime,m_{J/\psi}^2,m_{\omega}^2)}{
s^\prime-p^{\prime2}}\, , \nonumber\\
C^\prime_{\pi}&=&\int_{s^\prime_0}^{\infty}ds^\prime\frac{\tilde{\rho}_{\pi}(s^\prime,m_{\chi_{c1}}^2,m_{\pi}^2)}{
s^\prime-p^{\prime2}}\, , \nonumber\\
C^\prime_{D}&=&\int_{s^\prime_0}^{\infty}ds^\prime\frac{\tilde{\rho}_{D}(s^\prime,m_{D^*}^2,m_{D}^2)}{
s^\prime-p^{\prime2}}\, ,
\end{eqnarray}
where the densities  $\rho_{H}(s^\prime,s,u)=\tilde{\rho}_{\rho}(s^\prime,s,u)\delta(s-m_{J/\psi}^2)\delta(u-m_{\rho}^2)$, $\tilde{\rho}_{\omega}(s^\prime,s,u)\delta(s-m_{J/\psi}^2)\delta(u-m_{\omega}^2)$, $\tilde{\rho}_{\pi}(s^\prime,s,u)\delta(s-m_{\chi_{c1}}^2)\delta(u-m_{\pi}^2)$ and
$\tilde{\rho}_{D}(s^\prime,s,u)\delta(s-m_{D^*}^2)\delta(u-m_{D}^2)$, respectively. The densities $\tilde{\rho}_{\rho}(s^\prime,m_{J/\psi}^2,m_{\rho}^2)$, $\tilde{\rho}_{\omega}(s^\prime,m_{J/\psi}^2,m_{\omega}^2)$, $\tilde{\rho}_{\pi}(s^\prime,m_{\chi_{c1}}^2,m_{\pi}^2)$ and $\tilde{\rho}_{D}(s^\prime,m_{D^*}^2,m_{D}^2)$ are complex and we have no knowledge about the higher resonant states, as the spectrum is vague.
We  take the unknown functions $C^\prime_{\rho/\omega/\pi/D}$ as free parameters and adjust the suitable values to obtain flat Borel platforms for the hadronic coupling constants $G^\prime_{XJ/\psi\rho}$, $G^\prime_{XJ/\psi\omega}$, $G^\prime_{X\chi\pi}$ and $G^\prime_{XD^*D}$, respectively \cite{WZG-ZJX-Zc-Decay,WZG-Y4660-Decay}.

In Ref.\cite{X3872-Tetra-Nielsen-decay} (also Ref.\cite{Nielsen-review}), Navarra and Nielsen approximate  the hadron side of the correlation functions  as
\begin{eqnarray}
\Pi_{\rho}(p^{\prime2},p^2,q^2)&=& \frac{\lambda_X\, f_{J/\psi} m_{J/\psi}\,f_{\rho}m_{\rho}\,G_{XJ/\psi\rho} }{(m_{X}^2-p^{\prime2})(m_{J/\psi}^2-p^2)(m_{\rho}^2-q^2)} +\frac{B_{\rho}}{(s_0^\prime-p^{\prime2})(m_{\rho}^2-q^2)}\, , \nonumber\\
\Pi_{\omega}(p^{\prime2},p^2,q^2)&=& \frac{\lambda_X\, f_{J/\psi} m_{J/\psi}\,f_{\omega}m_{\omega}\,G_{XJ/\psi\omega} }{(m_{X}^2-p^{\prime2})(m_{J/\psi}^2-p^2)(m_{\omega}^2-q^2)} +\frac{B_{\omega}}{(s_0^\prime-p^{\prime2})(m_{\omega}^2-q^2)}\, ,
\end{eqnarray}
then only match them with the QCD side below the continuum threshold $s_0$, where the $B_{\rho/\omega}$ stand for the pole-continuum transitions, and we have changed  their notations (symbols) into the present form for
convenience. Although Navarra and Nielsen take account of the continuum contributions by introducing a parameter $s^\prime_0$ in the $s^\prime$ channel phenomenologically, they neglect the continuum contributions in the $u$ channel at the hadron side by hand. It is the shortcoming of that work. While in this work, we  match the hadron side with the QCD side  bellow the continuum thresholds $s_0$ and $u_0$ to obtain rigorous quark-hadron  duality, and we take account of the continuum contributions in the $s^\prime$ channel.

We set $p^{\prime2}=p^2$ in the correlation functions $\Pi_H(p^{\prime 2},p^2,q^2)$, and  perform  double Borel transform in regard  to  $P^2=-p^2$ and $Q^2=-q^2$, respectively, then we set $T_1^2=T_2^2=T^2$  to obtain  four QCD sum rules,
\begin{eqnarray} \label{X-rho-SR}
&&\frac{\lambda_{XJ/\psi\rho}G^\prime_{XJ/\psi\rho}}{m_{X}^2-m_{J/\psi}^2} \left[ \exp\left(-\frac{m_{J/\psi}^2}{T^2} \right)-\exp\left(-\frac{m_{X}^2}{T^2} \right)\right]\exp\left(-\frac{m_{\rho}^2}{T^2} \right)+C^\prime_{\rho} \exp\left(-\frac{m_{J/\psi}^2+m_{\rho}^2}{T^2}  \right) \nonumber\\
&&= \frac{\cos\theta-\sin\theta}{\sqrt{2}}\frac{m_c}{16\sqrt{2}\pi^4}\int_{4m_c^2}^{s_{J/\psi}^0}ds\int_{0}^{s^0_\rho}du\, \sqrt{1-\frac{4m_c^2}{s}}\exp\left(-\frac{s+u}{T^2}\right)\, ,
\end{eqnarray}

\begin{eqnarray} \label{X-omega-SR}
&&\frac{\lambda_{XJ/\psi\omega}G^\prime_{XJ/\psi\omega}}{m_{X}^2-m_{J/\psi}^2} \left[ \exp\left(-\frac{m_{J/\psi}^2}{T^2} \right)-\exp\left(-\frac{m_{X}^2}{T^2} \right)\right]\exp\left(-\frac{m_{\omega}^2}{T^2} \right)+C^\prime_{\omega} \exp\left(-\frac{m_{J/\psi}^2+m_{\omega}^2}{T^2}  \right) \nonumber\\
&&= \frac{\cos\theta+\sin\theta}{\sqrt{2}}\frac{m_c}{16\sqrt{2}\pi^4}\int_{4m_c^2}^{s_{J/\psi}^0}ds\int_{0}^{s^0_\omega}du\, \sqrt{1-\frac{4m_c^2}{s}}\exp\left(-\frac{s+u}{T^2}\right)\, ,
\end{eqnarray}

\begin{eqnarray} \label{X-pi-SR}
&&\frac{\lambda_{X\chi\pi}G^\prime_{X\chi\pi}}{m_{X}^2-m_{\chi_{c1}}^2} \left[ \exp\left(-\frac{m_{\chi_{c1}}^2}{T^2} \right)-\exp\left(-\frac{m_{X}^2}{T^2} \right)\right]\exp\left(-\frac{m_{\rho}^2}{T^2} \right)+C^\prime_{\pi} \exp\left(-\frac{m_{\chi_{c1}}^2+m_{\pi}^2}{T^2}  \right) \nonumber\\
&&= \frac{\cos\theta-\sin\theta}{\sqrt{2}}\frac{m_c\langle \bar{q}g_s\sigma G q\rangle}{16\sqrt{2}\pi^4}\int_{4m_c^2}^{s_{\chi_{c1}}^0}ds \, \frac{s-3m_c^2}{s\sqrt{s(s-4m_c^2)}}\exp\left(-\frac{s}{T^2}\right)\, ,
\end{eqnarray}

\begin{eqnarray} \label{X-D-SR}
&&\frac{\lambda_{XD^*D}G^\prime_{XD^*D}}{m_{X}^2-m_{D^*}^2} \left[ \exp\left(-\frac{m_{D^*}^2}{T^2} \right)-\exp\left(-\frac{m_{X}^2}{T^2} \right)\right]\exp\left(-\frac{m_{D}^2}{T^2} \right)+C^\prime_{D} \exp\left(-\frac{m_{D^*}^2+m_{D}^2}{T^2}  \right) \nonumber\\
&&= \cos\theta\frac{3m_c^2}{64\sqrt{2}\pi^4}\int_{m_c^2}^{s_{D^*}^0}ds\int_{m_c^2}^{s_{D}^0}du \,
\left( 1-\frac{m_c^2}{s}\right)^2\left( 1-\frac{m_c^2}{u}\right)^2\exp\left(-\frac{s+u}{T^2}\right)\nonumber\\
&&- \cos\theta\frac{m_c\langle\bar{q}q\rangle}{8\sqrt{2}\pi^2}\int_{m_c^2}^{s_{D^*}^0}ds \,
\left( 1-\frac{m_c^2}{s}\right)^2\exp\left(-\frac{s+m_c^2}{T^2}\right)\nonumber\\
&&- \cos\theta\frac{m_c\langle\bar{q}q\rangle}{8\sqrt{2}\pi^2}\int_{m_c^2}^{s_{D}^0}du \,
\left( 1-\frac{m_c^2}{u}\right)^2\exp\left(-\frac{u+m_c^2}{T^2}\right)\nonumber\\
&&+ \cos\theta\frac{m_c\langle\bar{q}g_s\sigma Gq\rangle}{32\sqrt{2}\pi^2T^2}\left(1+\frac{m_c^2}{T^2} \right)\int_{m_c^2}^{s_{D^*}^0}ds \,
\left( 1-\frac{m_c^2}{s}\right)^2\exp\left(-\frac{s+m_c^2}{T^2}\right)\nonumber\\
&&+ \cos\theta\frac{m_c\langle\bar{q}g_s\sigma Gq\rangle}{96\sqrt{2}\pi^2T^2}\left(1+\frac{3m_c^2}{T^2} \right)\int_{m_c^2}^{s_{D}^0}du \,
\left( 1-\frac{m_c^2}{u}\right)^2\exp\left(-\frac{u+m_c^2}{T^2}\right)\nonumber\\
&&- \cos\theta\frac{m_c\langle\bar{q}g_s\sigma Gq\rangle}{32\sqrt{2}\pi^2}\int_{m_c^2}^{s_{D^*}^0}ds \,
\frac{1}{s}\left( 1-\frac{m_c^2}{s}\right)\exp\left(-\frac{s+m_c^2}{T^2}\right)\nonumber\\
&&- \cos\theta\frac{m_c^3\langle\bar{q}g_s\sigma Gq\rangle}{96\sqrt{2}\pi^2}\int_{m_c^2}^{s_{D^*}^0}ds \,
\frac{1}{s^2}\exp\left(-\frac{s+m_c^2}{T^2}\right)\nonumber\\
&&- \cos\theta\frac{m_c\langle\bar{q}g_s\sigma Gq\rangle}{32\sqrt{2}\pi^2}\int_{m_c^2}^{s_{D}^0}du \,
\frac{1}{u}\left( 1-\frac{m_c^2}{u}\right)^2\exp\left(-\frac{u+m_c^2}{T^2}\right)\nonumber\\
&&+ \cos\theta\frac{m_c^3\langle\bar{q}g_s\sigma Gq\rangle}{96\sqrt{2}\pi^2}\int_{m_c^2}^{s_{D}^0}du \,
\frac{1}{u^2}\exp\left(-\frac{u+m_c^2}{T^2}\right)\, ,
\end{eqnarray}
where $\lambda_{XJ/\psi \rho}=\lambda_{X}\,f_{J/\psi}m_{J/\psi}\,f_{\rho}m_{\rho}$, $\lambda_{XJ/\psi \omega}=\lambda_{X}\,f_{J/\psi}m_{J/\psi}\,f_{\omega}m_{\omega}$, $\lambda_{X\chi\pi}=\lambda_{X}f_{\chi_{c1}}m_{\chi_{c1}}\mu_{\pi}$, and $\lambda_{XD^*D}=\lambda_{X}f_{D^*}m_{D^*}\mu_{D}$.

In Ref.\cite{X3872-Tetra-Nielsen-decay} (also Ref.\cite{Nielsen-review}), Navarra and Nielsen set  $p^{\prime2}=p^2$ in the correlation functions $\Pi_H(p^{\prime 2},p^2,q^2)$,  perform  single Borel transform in regard  to  $P^2=-p^2$, and take the $Q^2=-q^2$ as a free parameter to parameterize the off-shell-ness of the hadronic coupling constants $G_{XJ/\psi \rho}$ and $G_{XJ/\psi \omega}$, which are fitted into some functions of $Q^2$, then extract them to the physical points $q^2=m_{\rho/\omega}^2$, and finally  too large partial decay widths are obtained. The schemes are quite different,  we should not be surprised that the predictions in Ref.\cite{X3872-Tetra-Nielsen-decay} and in this  work are also quite different.

In calculations, we factorize out the mixing angle $\theta$ in Eqs.\eqref{X-rho-SR}-\eqref{X-D-SR} so as to facilitate  determining  the mixing effects, and redefine the hadronic coupling constants $G$ and free parameters $C$,
\begin{eqnarray}
G^\prime_{XJ/\psi\rho}&=&G_{XJ/\psi\rho}\frac{\cos\theta-\sin\theta}{\sqrt{2}}\, ,\nonumber\\
G^\prime_{XJ/\psi\omega}&=&G_{XJ/\psi\omega}\frac{\cos\theta+\sin\theta}{\sqrt{2}}\, ,\nonumber\\
G^\prime_{X\chi\pi}&=&G_{X\chi\pi}\frac{\cos\theta-\sin\theta}{\sqrt{2}}\, ,\nonumber\\
G^\prime_{XD^*D}&=&G_{XD^*D}\cos\theta\, ,
\end{eqnarray}
\begin{eqnarray}
C^\prime_{\rho}&=&C_{\rho}\frac{\cos\theta-\sin\theta}{\sqrt{2}}\, ,\nonumber\\
C^\prime_{\omega}&=&C_{\omega}\frac{\cos\theta+\sin\theta}{\sqrt{2}}\, ,\nonumber\\
C^\prime_{\pi}&=&C_{\pi}\frac{\cos\theta-\sin\theta}{\sqrt{2}}\, ,\nonumber\\
C^\prime_{D}&=&C_{D}\cos\theta\, ,
\end{eqnarray}
then it is easy to study the dependence on the mixing angle $\theta$.

\section{Numerical results and discussions}
We take  the conventional  vacuum condensates,
$\langle \bar{q}q \rangle=-(0.24\pm 0.01\, \rm{GeV})^3$,
$\langle\bar{q}g_s\sigma G q \rangle=m_0^2\langle \bar{q}q \rangle$,
$m_0^2=(0.8 \pm 0.1)\,\rm{GeV}^2$     at the   energy scale  $\mu=1\, \rm{GeV}$
\cite{SVZ79,Reinders85,Colangelo-Review},  and take the $\overline{MS}$  mass $m_{c}(m_c)=(1.275\pm0.025)\,\rm{GeV}$ from the Particle Data Group \cite{PDG}. We set $m_u=m_d=0$ and take account of
the energy-scale dependence from re-normalization group equation,
\begin{eqnarray}
\langle\bar{q}q \rangle(\mu)&=&\langle\bar{q}q \rangle({\rm 1GeV})\left[\frac{\alpha_{s}({\rm 1GeV})}{\alpha_{s}(\mu)}\right]^{\frac{12}{33-2n_f}}\, , \nonumber\\
 \langle\bar{q}g_s \sigma Gq \rangle(\mu)&=&\langle\bar{q}g_s \sigma Gq \rangle({\rm 1GeV})\left[\frac{\alpha_{s}({\rm 1GeV})}{\alpha_{s}(\mu)}\right]^{\frac{2}{33-2n_f}}\, , \nonumber\\
 m_c(\mu)&=&m_c(m_c)\left[\frac{\alpha_{s}(\mu)}{\alpha_{s}(m_c)}\right]^{\frac{12}{33-2n_f}} \, ,\nonumber\\
 \alpha_s(\mu)&=&\frac{1}{b_0t}\left[1-\frac{b_1}{b_0^2}\frac{\log t}{t} +\frac{b_1^2(\log^2{t}-\log{t}-1)+b_0b_2}{b_0^4t^2}\right]\, ,
\end{eqnarray}
  where   $t=\log \frac{\mu^2}{\Lambda_{QCD}^2}$, $b_0=\frac{33-2n_f}{12\pi}$, $b_1=\frac{153-19n_f}{24\pi^2}$, $b_2=\frac{2857-\frac{5033}{9}n_f+\frac{325}{27}n_f^2}{128\pi^3}$,  $\Lambda_{QCD}=210\,\rm{MeV}$, $292\,\rm{MeV}$  and  $332\,\rm{MeV}$ for the flavors  $n_f=5$, $4$ and $3$, respectively  \cite{PDG,Narison-mix}, and we choose  $n_f=4$, and evolve all the input parameters to the energy scale  $\mu=1\,\rm{GeV}$.

 At the hadron side, we take $m_{\pi^\pm}=0.13957\,\rm{GeV}$, $m_{\pi^0}=0.13498\,\rm{GeV}$, $m_{J/\psi}=3.0969\,\rm{GeV}$, $m_{\chi_{c1}}=3.51067\,\rm{GeV}$, $m_\rho=0.77526\,\rm{GeV}$, $m_\omega=0.78266\,\rm{GeV}$,
  $f_{\pi}=0.130\,\rm{GeV}$ from the Particle Data Group \cite{PDG},
 $m_{D^*}=2.01\,\rm{GeV}$, $m_{D}=1.87\,\rm{GeV}$, $f_{D^*}=263\,\rm{MeV}$, $f_{D}=208\,\rm{MeV}$, $s^0_{D^*}=6.4\,\rm{GeV}^2$, $s^0_{D}=6.2\,\rm{GeV}^2$   \cite{WangJHEP}, $f_{J/\psi}=0.418 \,\rm{GeV}$  \cite{Becirevic}, $f_{\chi_{c1}}=0.338\,\rm{GeV}$ \cite{Charmonium-PRT}, $f_{\rho}=0.215\,\rm{GeV}$, $f_{\omega}=f_{\rho}$,  $\sqrt{s^0_{\rho}}=1.2\,\rm{GeV}$, $s^0_\omega=s^0_{\rho}$  \cite{PBall-decay-Kv},
 $m_{X}=3.91\,\rm{GeV}$,   $\lambda_{X}=2.10 \times 10^{-2}\,\rm{GeV}^5$ \cite{WZG-PRD-cc-spectrum} from the QCD sum rules, and  $f_{\pi}m^2_{\pi}/(m_u+m_d)=-2\langle \bar{q}q\rangle/f_{\pi}$ from the Gell-Mann-Oakes-Renner relation.

\begin{figure}
 \centering
  \includegraphics[totalheight=5cm,width=7cm]{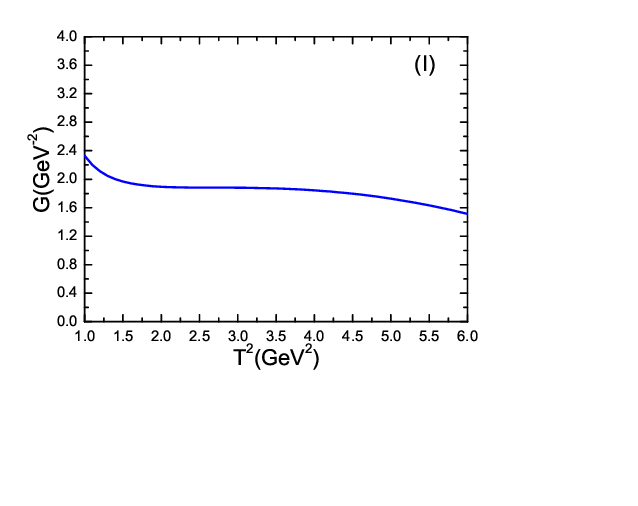}
  \includegraphics[totalheight=5cm,width=7cm]{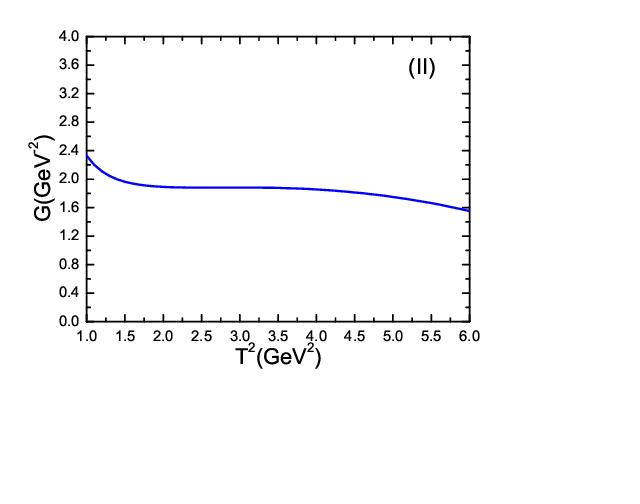}
   \includegraphics[totalheight=5cm,width=7cm]{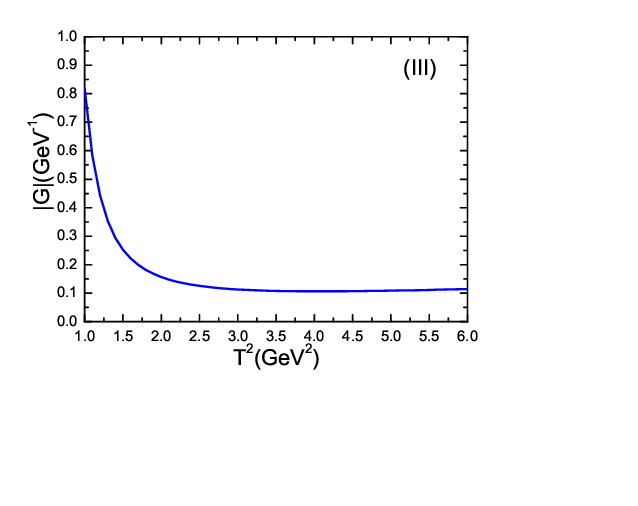}
  \includegraphics[totalheight=5cm,width=7cm]{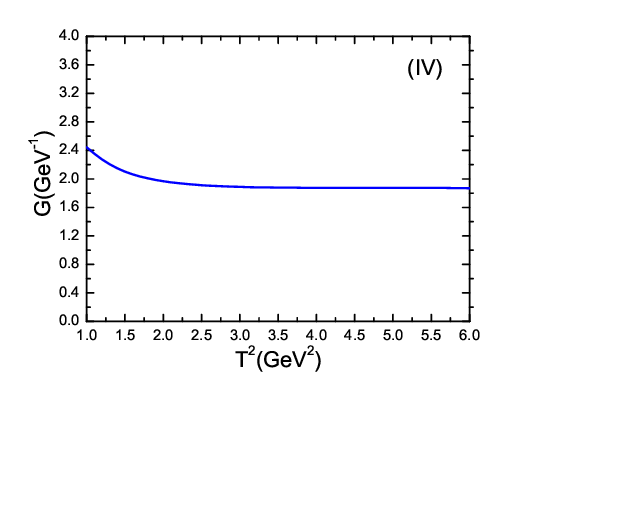}
     \caption{The central values of the hadronic coupling constants  with variations of the  Borel  parameters  $T^2$, where the (I), (II), (III) and (IV)  denote the $G_{XJ/\psi \rho}$,
     $G_{XJ/\psi \omega}$, $G_{X\chi \pi}$ and $G_{XD^*D}$,  respectively.}\label{hadron-coupling-fig}
\end{figure}

In calculations, we fit the free parameters as  $C_{\rho}=0.000250(T^2-1.5\,\rm{GeV}^2)\,\rm{GeV}^3$,
$C_{\omega}=0.000245(T^2-1.5\,\rm{GeV}^2)\,\rm{GeV}^3$, $C_{\pi}=0$ and $C_{D}=0.0000725(T^2-2.1\,\rm{GeV}^2)\,\rm{GeV}^4$ to obtain the Borel windows  $T^2_{\rho}=(2.3-3.3)\,\rm{GeV}^2$, $T^2_{\omega}=(2.3-3.3)\,\rm{GeV}^2$,
$T^2_{\pi}=(3.6-4.6)\,\rm{GeV}^2$ and $T^2_{D}=(4.0-5.0)\,\rm{GeV}^2$,
where  the subscripts $\rho$, $\omega$, $\pi$ and $D$ denote the corresponding channels.
We obtain uniform enough flat platforms  $T^2_{max}-T^2_{min}=1\,\rm{GeV}^2$, where the max and min denote the maximum and minimum, respectively.
 In Fig.\ref{hadron-coupling-fig}, we plot the hadronic coupling constants $G_{XJ/\psi \rho}$,
     $G_{XJ/\psi \omega}$, $G_{X\chi \pi}$ and $G_{XD^*D}$ with variations of the Borel parameters at large intervals. In the Borel windows, there appear very flat platforms indeed, it is reliable to extract the hadron coupling constants.

Now, we estimate the uncertainties in the following ways. For example, the  uncertainties of an input parameter $\xi$, $\xi= \bar{\xi} +\delta \xi$,  result in the uncertainties $\lambda_{X}f_{J/\psi}f_{\rho}G_{XJ/\psi \rho} = \bar{\lambda}_{X}\bar{f}_{J/\psi}\bar{f}_{\rho}\bar{G}_{XJ/\psi \rho}
+\delta\,\lambda_{X}f_{J/\psi}f_{\rho}G_{XJ/\psi \rho}$, $C_{\rho} = \bar{C}_{\rho}+\delta C_{\rho}$,
\begin{eqnarray}\label{Uncertainty-4}
\delta\,\lambda_{X}f_{J/\psi}f_{\rho}G_{XJ/\psi \rho} &=&\bar{\lambda}_{X}\bar{f}_{J/\psi}\bar{f}_{\rho}\bar{G}_{XJ/\psi \rho}\left( \frac{\delta f_{J/\psi}}{\bar{f}_{J/\psi}} +\frac{\delta f_{\rho}}{\bar{f}_{\rho}}+\frac{\delta \lambda_{X}}{\bar{\lambda}_{X}}+\frac{\delta G_{XJ/\psi \rho}}{\bar{G}_{XJ/\psi \rho}}\right)\, ,
\end{eqnarray}
we can set $\delta C_{\rho}=0$ and $ \frac{\delta f_{J/\psi}}{\bar{f}_{J/\psi}} =\frac{\delta f_{\rho}}{\bar{f}_{\rho}}=\frac{\delta \lambda_{X}}{\bar{\lambda}_{X}}=\frac{\delta G_{XJ/\psi \rho}}{\bar{G}_{XJ/\psi \rho}}$ approximately.

Finally, we obtain the values of the hadronic coupling constants,
\begin{eqnarray} \label{HCC-values}
G_{XJ/\psi\rho} &=&1.88^{+0.11}_{-0.10}\,\rm{GeV}^{-2}\, , \nonumber\\
G_{XJ/\psi\omega} &=&1.88^{+0.11}_{-0.10}\,\rm{GeV}^{-2}\, , \nonumber\\
|G_{X\chi\pi}| &=&0.11 \,\rm{GeV}^{-1} \, , \nonumber\\
G_{XD^*D} &=&1.875^{+0.064}_{-0.064}\,\rm{GeV}^{-1}\, ,
\end{eqnarray}
where we only present the central value of the $G_{X\chi\pi}$ due to the tiny partial decay width of the $X(3872)\to \chi_{c1}\pi^0$.

Now we take the hadron masses $m_{X}=3.87165\,\rm{GeV}$,  $m_{D^{*0}}= 2.00685\,\rm{GeV}$, $m_{D^{0}}= 1.86484\,\rm{GeV}$ and
$m_{J/\psi}= 3.09690\,\rm{GeV}$ from the Particle Data Group to calculate the partial decay widths \cite{PDG}.
As the $X(3872)$ lies near the thresholds of the final states $J/\psi\rho$, $J/\psi\omega$ and $D^*\bar{D}$, we should take account of the finite width effects of the $\rho$, $\omega$ and $D^*$ mesons, due to the decay cascades,
\begin{eqnarray}
X(3872)&\to&J/\psi \rho^0 \to J/\psi \pi^+\pi^-\, , \nonumber \\
X(3872)&\to&J/\psi \omega\to J/\psi \pi^+\pi^-\pi^0\, , \nonumber \\
X(3872)&\to&D^{*0} \bar{D}^0\to D^0\bar{D}^0 \pi^0\, .
\end{eqnarray}
Then we obtain the partial decay widths via trial and error, as there is an additional parameter $\theta$, $\theta=0.12\pi=21.6^\circ$. The
partial widths are listed in the following,
\begin{eqnarray} \label{Par-width-rho}
\Gamma(X\to J/\psi \pi\pi)&=&\frac{1
}{24\pi^2 m_{X}^2  } \int_{\Delta_{2\pi}^2}^{(m_{X}-m_{J/\psi})^2}ds\,|T_{\rho}|^2\frac{m_\rho\Gamma_{\rho}\,
p(m_{X},m_{J/\psi},\sqrt{s})}{(s-m_{\rho}^2)^2+m_\rho^2\Gamma_{\rho}^2}\ , \nonumber\\
&=& 0.132^{+0.016}_{-0.013} \,\rm{MeV}\, ,
\end{eqnarray}

\begin{eqnarray} \label{Par-width-omega}
\Gamma(X\to J/\psi \pi\pi\pi)&=&\frac{1
}{24\pi^2 m_{X}^2  } \int_{\Delta_{3\pi}^2}^{(m_{X}-m_{J/\psi})^2}ds\,|T_{\omega}|^2\frac{m_\rho\Gamma_{\rho}\,
p(m_{X},m_{J/\psi},\sqrt{s})\,{\rm Br}}{(s-m_{\omega}^2)^2+m_\omega^2\Gamma_{\omega}^2}\ , \nonumber\\
&=& 0.129^{+0.016}_{-0.013} \,\rm{MeV}\, ,
\end{eqnarray}

\begin{eqnarray} \label{Par-width-pi}
\Gamma(X\to \chi_{c1} \pi)&=&\frac{|T_{\pi}|^2\,p(m_{X},m_{\chi_{c1}},m_\pi)
}{24\pi m_{X}^2} \ , \nonumber\\
&=& 0.0016 \,\rm{MeV}\, ,
\end{eqnarray}

\begin{eqnarray} \label{Par-width-D}
\Gamma(X\to D^0\bar{D}^0 \pi^0)&=&\frac{1
}{24\pi^2 m_{X}^2  } \int_{(m_D+m_\pi)^2}^{(m_{X}-m_{D})^2}ds\,|T_{D}|^2\frac{m_{D^*}\Gamma_{D^*}\,
p(m_{X},m_{D},\sqrt{s})}{(s-m_{D^*}^2)^2+m_{D^*}^2\Gamma_{D^*}^2}\ , \nonumber\\
&=& 2.262^{+0.157}_{-0.152} \,\rm{MeV}\,\,\, {\rm for}\,\,\, \Gamma_{D^{*0}}=2.0\,\rm{MeV} \, ,\nonumber\\
&=& 1.795^{+0.124}_{-0.121} \,\rm{MeV}\,\,\, {\rm for}\,\,\, \Gamma_{D^{*0}}=1.0\,\rm{MeV} \, ,\nonumber\\
&=& 1.326^{+0.092}_{-0.089} \,\rm{MeV}\,\,\, {\rm for}\,\,\, \Gamma_{D^{*0}}=0.5\,\rm{MeV} \, ,\nonumber\\
&=& 0.485^{+0.034}_{-0.032} \,\rm{MeV}\,\,\, {\rm for}\,\,\, \Gamma_{D^{*0}}=0.1\,\rm{MeV} \, ,
\end{eqnarray}
where
\begin{eqnarray}
|T_\rho|^2&=& G_{XJ/\psi\rho}^{\prime2} \frac{\left(m_{X}^2-m_{J/\psi}^2-s\right)^2}{4}\left[\frac{\left(m_{X}^2-s\right)^2}{2m_{J/\psi}^2}
+\frac{\left(m_{X}^2-m_{J/\psi}^2\right)^2}{2s}  +4m_X^2-\frac{m_{J/\psi}^2+s}{2}\right] \, ,\nonumber
\end{eqnarray}

\begin{eqnarray}
|T_\omega|^2&=& G_{XJ/\psi\omega}^{\prime2} \frac{\left(m_{X}^2-m_{J/\psi}^2-s\right)^2}{4}\left[\frac{\left(m_{X}^2-s\right)^2}{2m_{J/\psi}^2}
+\frac{\left(m_{X}^2-m_{J/\psi}^2\right)^2}{2s}  +4m_X^2-\frac{m_{J/\psi}^2+s}{2}\right] \, ,\nonumber
\end{eqnarray}

\begin{eqnarray}
|T_\pi|^2&=& G_{X\chi\pi}^{\prime2}  \frac{\lambda(m_X^2,m_{\chi_{c1}}^2,m_{\pi}^2)}{2} \, ,\nonumber
\end{eqnarray}

\begin{eqnarray}
|T_D|^2&=& G_{XD^*D}^{\prime2} \frac{\left(m_{X}^2-m_{D}^2-s\right)^2}{4}\left[3+\frac{\lambda(m_X^2,s,m_{D}^2)}{4m_X^2s}\right] \, ,\nonumber
\end{eqnarray}

\begin{eqnarray}
G^\prime_{XJ/\psi\rho}&\to&\widetilde{G}^\prime_{XJ/\psi\rho}=G^\prime_{XJ/\psi\rho}\exp\left(
-\frac{m_\rho^2-s}{s-\Delta_{2\pi}^2} \frac{m_\rho^4}{\Delta_{2\pi}^4}\right) \, , \nonumber\\
G^\prime_{XJ/\psi\omega}&\to&\widetilde{G}^\prime_{XJ/\psi\omega}=G^\prime_{XJ/\psi\omega}\exp\left(-\frac{m_\omega^2-s}{s-\Delta_{3\pi}^2} \frac{m_\omega^6}{\Delta_{3\pi}^6}\right) \, ,
\end{eqnarray}
$\Delta_{2\pi}=m_{\pi^+}+m_{\pi^-}$, $\Delta_{3\pi}=m_{\pi^+}+m_{\pi^-}+m_{\pi^0}$,  ${\rm Br (\omega \to \pi\pi\pi)}=0.892$ \cite{PDG}, $\lambda(a,b,c)=a^2+b^2+c^2-2ab-2ac-2bc$, $p(A,B,C)=\frac{\sqrt{\left[A^2-(B+C)^2\right]\left[A^2-(B-C)^2\right]}}{2A}$.
The hadronic coupling constants from the QCD sum rules in Eqs.\eqref{X-rho-SR}-\eqref{X-D-SR} are physical
quantities under zero width approximation. The physical widths from the Particle Data Group are $\Gamma_\rho=147.4\,\rm{MeV}$, $\Gamma_\omega=8.68\,\rm{MeV}$, $\Gamma_{\chi_{c1}}=0.88\,\rm{MeV}$,
$\Gamma_{D^{*0}}< 2.1\,\rm{MeV}$, respectively, we introduce  exponential form-factors $\exp\left(
-\frac{m_\rho^2-s}{s-\Delta_{2\pi}^2} \frac{m_\rho^4}{\Delta_{2\pi}^4}\right)$ and
$ \exp\left(-\frac{m_\omega^2-s}{s-\Delta_{3\pi}^2} \frac{m_\omega^6}{\Delta_{3\pi}^6}\right)$ to parameterize the "off-shell" effects due to the $J/\psi\rho$ and $J/\psi\omega$ thresholds, as the $X(3872)$ lies near the $J/\psi\rho$ and $J/\psi\omega$ thresholds. At the mass-shell $s=m_\rho^2$ and $m_\omega^2$, they reduce to 1 to match with the zero width approximation in the QCD sum rules. At the thresholds, $s=\Delta_{2\pi}^2$ and $\Delta_{3\pi}^2$,  the available phase-spaces are very small, the decays $\rho \to \pi\pi$ and $\omega\to\pi\pi\pi$ only take place through the lower tails, which can be taken as some intermediate sates with the same quantum numbers as the $\rho$ and $\omega$ except for the masses,  and  are greatly suppressed.  On the other hand, the "off-shell" effects on the hadronic coupling constants are considerable, we should introduce some form-factors to parameterize them.

The width $\Gamma_{D^{*\pm}}=0.0834\pm0.0018\,\rm{MeV}$ from the Particle Data Group \cite{PDG}, if we take the approximation $\Gamma_{D^{*0}}\approx\Gamma_{D^{*\pm}}\approx 0.1\,\rm{MeV} $, then
$\Gamma(X\to D^{*0}\bar{D}^0)$+$\Gamma(X\to \bar{D}^{*0}D^0)=0.970^{+0.068}_{-0.064} \,\rm{MeV}$
from Eq.\eqref{Par-width-D}, which is in excellent agreement  with the branching fraction $(52.4^{+25.3}_{-14.3}) \%$ from the combined data analysis \cite{X3872-YuanCZ} and the total width $ \Gamma_{X}=1.19\pm 0.21\,\rm{MeV}$ from the Particle Data Group \cite{PDG}. On the other hand, if we take the masses of the $X(3872)$, $D^{\pm}$, $D^0$, $\pi^\pm$ and $\pi^0$ from the Particle Data Group \cite{PDG}, the decays $X(3872)\to D^{*+}\bar{D}^-\to \bar{D}^+\bar{D}^-\pi^0$ and $X(3872)\to D^{*+}\bar{D}^-\to \bar{D}^0\bar{D}^-\pi^+$ cannot take place due to the negative phase-space.

The partial widths $\Gamma(X\to J/\psi \pi\pi)= 0.132^{+0.016}_{-0.013} \,\rm{MeV}$ and $\Gamma(X\to J/\psi \pi\pi\pi)= 0.129^{+0.016}_{-0.013} \,\rm{MeV}$  from Eq.\eqref{Par-width-rho} and Eq.\eqref{Par-width-omega}
are in excellent agreement with the branching fractions ${\rm Br} (X\to J/\psi \pi\pi)=(4.1^{+1.9}_{-1.1})\%$ and  ${\rm Br} (X\to J/\psi \omega)=(4.4^{+2.3}_{-1.3})\%$ from the combined data analysis \cite{X3872-YuanCZ}, so the mixing angle $\theta=21.6^\circ$, which is compatible with the values $\theta=20.0^\circ$ \cite{X3872-Tetra-Maiani} and $\theta=23.5^\circ$ \cite{X3872-Tetra-Nielsen-decay}. The significant difference is that we take account of all the Feynman diagrams and take rigorous quark-hadron duality, while in Ref.\cite{X3872-Tetra-Nielsen-decay}, only the connected diagrams are taken into account to obtain small partial decay widths.

The partial decay width $\Gamma(X\to \chi_{c1} \pi)=0.0016 \,\rm{MeV}$ from Eq.\eqref{Par-width-pi}
is much smaller than the branching fraction   $(3.6^{+2.2}_{-1.6})\%$ from the combined data analysis \cite{X3872-YuanCZ} or the branching fraction   $(3.4\pm1.6)\%$ from the Particle Data Group \cite{PDG}, more precise measurement is still needed.

All in all, in this work, we reproduce the small width of the $X(3872)$ via the QCD sum rules for the first time.

\section{Conclusion}
 In this work, we take the $X(3872)$ as the hidden-charm tetraquark state with both isospin $I=0$ and $1$ components, then investigate the hadronic coupling constants $G^\prime_{XJ/\psi \rho}$,
     $G^\prime_{XJ/\psi \omega}$, $G^\prime_{X\chi \pi}$ and $G^\prime_{XD^*D}$ with the QCD sum rules in details. We select the optimal tensor structures and take account of all the Feynman diagrams, then acquire
     four QCD sum rules based on the  rigorous quark-hadron duality. After careful calculations, we obtain the hadronic coupling constants, then determine the mixing angle via trial and error, and obtain the partial decay widths for the $X(3872)\to J/\psi \pi^+\pi^-$, $J/\psi\omega$,  $\chi_{c1}\pi^0$, $D^{*0}\bar{D}^0$ and $D^{0}\bar{D}^0\pi^0$. The total width is about $1\,\rm{MeV}$, which is in excellent agreement with the experiment data $ \Gamma_{X}=1.19\pm 0.21\,\rm{MeV}$  from the Particle Data Group, it is  the first time  to reproduce the small width of the $X(3872)$ via the QCD sum rules. The present calculations support assigning the $X(3872)$ as the mixed hidden-charm tetraquark state with the quantum numbers $J^{PC}=1^{++}$.

\section*{Acknowledgements}
This  work is supported by National Natural Science Foundation, Grant Number  12175068.

\end{document}